\documentclass[preprint,aps]{revtex4}
\usepackage[dvips]{graphicx}
\usepackage{amsmath,accents}
\usepackage{amssymb}
\DeclareMathSymbol{\widehatsym}{\mathord}{largesymbols}{"65}
\newcommand\lowerwidehatsym{%
  \text{\smash{\raisebox{-1.3ex}{%
    $\widehatsym$}}}}
\newcommand\wtil[1]{%
  \mathchoice
    {\accentset{\displaystyle\lowerwidehatsym}{#1}}
    {\accentset{\textstyle\lowerwidehatsym}{#1}}
    {\accentset{\scriptstyle\lowerwidehatsym}{#1}}
    {\accentset{\scriptscriptstyle\lowerwidehatsym}{#1}}
}
\draft
\begin{document}
\title{Soliton solution of the Zakharov equations in quantum plasmas}
\author{F. Sayed$^{1}$, S. V. Vladimirov$^{1,2,3}$, Yu. Tyshetskiy$^{1}$, and O. Ishihara$^{2}$}
\address{$^{1}$School of Physics, University of Sydney, New South Wales 2006, Australia\\
$^{2}$Faculty of Engineering, Yokohama National University, Yokohama 240-8501, Japan\\
$^{3}$Metamaterials Laboratory, National Research University of Information Technology,
Mechanics, and Optics, St Petersburg 199034, Russia
}
\begin{abstract}
We investigate the existence of envelope soliton solutions 
in collisionless quantum plasmas, using the quantum-corrected Zakharov equations 
in the kinetic case, which describes the interaction between high frequency 
Langmuir waves and low frequency plasma density variations. We show the role 
played by quantum effects in the nonlinearity/dispersion balance leading to the 
formation of soliton solutions of the 
quantum-corrected nonlinear Schrodinger (QNLS) equation.
\end{abstract}
\maketitle
\begin{sloppypar}
\section{Introduction}
A soliton is a special type of solitary wave (a hump- or dip - shaped nonlinear wave) which preserves its shape and speed, even after collisions with other solitary wave. It arises because of the balance
between the effects of nonlinearity and dispersion (when the effect of dissipation is
negligible in comparison with those of the nonlinearity and dispersion). One-dimensional solitons
into plasma physics were first considered in 1963 by Gurevich
and Pitaevskii~\cite{r1}. Modulational instabilities are responsible 
for the formation of Langmuir envelope 
solitons (cavitons)~\cite{r2,r3,r4,r5,r6}. Electromagnetic 
envelope solitons were investigated for the first time by Hasegawa~\cite{r7} 
and Karpman~\cite{r8}. Currently, many types of solitons in
plasmas are known, e. g. one-dimensional solitons and two or three-dimensional solitons~\cite{r9}.
They are invoked in various physical theories, especially for the
construction of different versions of strong turbulence theory. In order for solitons to be applicable in these theories it is necessary that they be stable. Therefore, the problem of soliton stability is of particular importance~\cite{r10}.

Quantum plasmas are ubiquitous and appear in different physical settings from laser-matter interaction experiments (e.g., the compressed hydrogen in the fast ignition scenario of inertial fusion is in a quantum 
plasma state), to astrophysical and cosmological
objects~\cite{r11,r12,r13} (e.g., interstellar or molecular clouds,
planetary rings, comets, interiors of white dwarf stars, etc.),
nanostructures~\cite{r14}, and microelectronic devices~\cite{r15}.
At room temperature and standard metallic
densities, the electron gas in an ordinary metal is a good
example of a quantum plasma system. In such plasmas, quantum mechanical effects
(e.g., tunneling) are important since the de Broglie wavelength
of the charge carriers (e.g., electrons and holes or positrons) is comparable 
to the dimension of the system.
Recently, the topic of quantum plasmas has attracted considerable 
attention~\cite{r16,r17,r18,r19,r20,r21,r22,r23} and it is desirable that a 
good understanding of the basic phenomena of quantum plasmas and be achieved.

In classical plasmas, nonlinear phenomena are often formulated in terms of completely
integrable evolution equations of the Korteweg-de Vries (KdV) or nonlinear Schrodinger equation
(NLS) type~\cite{r24,r25,r26}. These completely integrable equations, as is well known, admit N-soliton solutions, derivable from the Inverse Scattering Transform (IST) method.

In plasma, large amplitude Langmuir waves can initiate a number of nonlinear effects including decay and modulational instabilities, which are described by the Zakharov equations~\cite{r26,r27}. The Zakharov equations in a quantum plasma (which we hereby call the quantum-corrected Zakharov equations) were recently obtained in the electrostatic approximation using the Wigner kinetic approach for the electron quantum distribution function~\cite{r28}. It was shown in~\cite{r28} that in quantum plasma, the classical Zakharov equations are modified by quantum correction terms proportional to $\hbar^2$ (note that in the formal classical limit $\hbar=0$, the quantum-corrected Zakharov equations obtained in~\cite{r28} reduce to the classical Zakharov equations~\cite{r6}).

Nonlinear waves in quantum plasmas are
not yet sufficiently well understood. For example, the classical Zakharov 
equations in the adiabatic limit reduce to the nonlinear Schrodinger (NLS)
equation which is completely integrable via the IST (inverse scattering
transform) method, but no such technique seems to be easily constructable in the 
quantum case.

In terms of the variational 
formulation of quantum-modified NLS equation, Haas et al.~\cite{r29} investigated the coefficients of soliton solution in quantum-corrected form (when $v_0=0$, where $v_0$ is a constant speed at which the soliton solution propagates).
In the present work, we will focus on the soliton solution of the quantum-corrected Zakharov equations which propagates with constant speed $v_0$. The purpose of this study is to investigate the role played by quantum effects in the quantum-corrected Zakharov system; these effects modify the dispersion-nonlinearity equilibrium, which is the ultimate factor responsible for the existence of solitons.

In this paper, we investigate the previously obtained quantum-corrected Zakharov equations relevant to quantum plasmas, and obtain their quantum-corrected soliton solutions, comparing them with those in the case of the classical Zakharov model.

\section{Soliton solution of the quantum Zakharov equations}
In the previous work~\cite{r28}, we have generalized the formalism of modulational interactions to nonrelativistic quantum plasmas, based on the Wigner kinetic description of collisionless quantum plasmas.
In particular, we derived kinetically the effective cubic response of a quantum plasma 
(which in general is a complex-valued function), which can be used for various modulational 
processes. We derived the quantum-corrected Zakharov equations 
for collisionless quantum plasmas by neglecting the imaginary part of the effective cubic 
response. The resulting quantum-corrected Zakharov equations describe the coupled nonlinear 
evolution of high-frequency fields and low-frequency density perturbations in collisionless 
quantum plasmas that can be written in the one dimensional case as
\begin{eqnarray}
&&\hspace{-1.8cm}\left(i \frac{\partial}{\partial
t}+\frac{3}{2}\frac{v^2_e}{\omega_{pe}}\frac{\partial^2}{\partial x^2}
-\frac{\hbar^2}{8m^2_e\omega_{pe}}\frac{\partial^4}{\partial
x^4}\right)E(x,t) =\frac{\omega_{pe}}{2}\frac{\delta n
(x,t)}{n_0}E(x,t),\label{q1}
\end{eqnarray}
\begin{eqnarray}
&&\hspace{-1.8cm}\left(\frac{\partial^2}{\partial
t^2}+\frac{\hbar^2}{12m^2_ev^2_e}\frac{\partial^4}{\partial t^2
\partial x^2}
-v^2_s\frac{\partial^2}{\partial x^2}\right)\frac{\delta n
(x,t) }{n_0} =\left(\frac{\partial^2}{\partial
x^2}+\frac{\hbar^2}{12m^2_ev^2_e}\frac{\partial^4}{\partial x^4}
\right)\frac{|E(x,t)|^2}{4\pi n_0 m_i}, \label{q2}
\end{eqnarray}
where $E(x,t)$ is the envelope of the Langmuir wave packet modulated by the nonlinear interaction with plasma density variations, $\delta n(x,t)$ is the plasma density variations from its equilibrium value $n_0$, $\omega_{pe}$ is the electron plasma frequency (i.e. $\omega_{pe}=\sqrt{{4\pi e^2n_0/m_e}}$), $m_e$ and $m_i$ are the electron and ion masses, respectively, $v_e$ is the electrons thermal velocity and $v_s$ 
is the ion sound velocity, and $\hbar$ is the reduced Planck constant.

Upon introducing the dimensionless variables
\begin{eqnarray}
&&\hspace{-1.8cm}{
x}=\frac{2}{3}\frac{m_e}{m_i}\frac{\omega_{pe}}{v_s}{X},~~~~~t= \frac{2}{3}\frac{m_e}{m_i}\omega_{pe}\tau,\nonumber\\
&&\hspace{-3.4cm} ~~~~~~{\it E}=\sqrt{16\pi
n_0T_em_e/3m_i}\wtil{E},~~~\frac{\delta n}{n_0}=\frac{4}{3}\frac{m_e}{m_i}\wtil{\delta n},\nonumber\\
&&\hspace{-2.7cm}~~~~~~H=\frac{\hbar\omega_{pe}}{m_ev^2_s
},\label{q3}
\end{eqnarray}
the corresponding dimensionless Zakharov equations in quantum plasma can be presented in the form
\begin{eqnarray}
&&\hspace{-3.8cm}\left(i \frac{\partial}{\partial
\tau}+\frac{\partial^2}{\partial X^2}-H^2\frac{\partial^4}{\partial
X^4}\right)\wtil{E}=\wtil{\delta n} \wtil{E},\label{q4}
\end{eqnarray}
\begin{eqnarray}
&&\hspace{-2.8cm}\left(\frac{\partial^2}{\partial
\tau^2}+H^2\frac{\partial^4}{\partial \tau^2
\partial X^2}
-\frac{\partial^2}{\partial X^2}\right)\wtil{\delta n}
=\frac{\partial^2}{\partial
X^2}\left(1+H^2\frac{\partial^2}{\partial X^2}\right){|
\wtil{E}|^2},\label{q5}
\end{eqnarray}
where the quantum parameter $H$ is given in Eq.~(\ref{q3}). In dense plasmas~\cite{r11,r13}, the particle density is about $10^{25}$ -- $10^{32}$ $m^{-3}$ and temperature is about $10^5$ -- $10^7$ K. For a completely ionized hydrogen plasma in these density and temperature ranges, $H$ typically ranges from about $10^{-5}$ up to values of order unity.
For large values of $H$ (i. e. $H\sim1$) particularly in astrophysical plasmas, quantum effects in the coupling between Langmuir and ion-acoustic modes become important~\cite{r30}.

If we ignore the quantum correction by setting $H=0$, we simply obtain the 
classical Zakharov equations~\cite{r6}. At the classical level, a set of coupled nonlinear wave equations describing the interaction between high-frequency Langmuir waves with low frequency 
plasma density variations was first derived by Zakharov~\cite{r26,r27}. For the classical 
model, one can find many kinds of solitons by various methods~\cite{r31,r32,r33}. In 
the quantum case, the analysis of existence of localized or soliton solutions is much more  difficult. The basic difficulty follows from the fact that the new, quantum-corrected equations constitute a more complicated  system of coupled, fourth-order nonlinear equations~(\ref{q4})-(\ref{q5}).

In the adiabatic limit, by setting $[(\partial^2 /\partial \tau^2)(\wtil{\delta n})]=0$ in Eq.~(\ref{q5}), solutions of the classical ($H=0$) Zakharov system are found.
In this situation, the envelope of the electric field satisfies a nonlinear 
Schrodinger equation, which is completely integrable yielding N-soliton solutions~\cite{r6}. 
Solitons usually arise as a consequence of the detailed balance 
between dispersive and nonlinear contributions. The quantum effects may perturb or 
perhaps even destroy these localized solitonic solutions. Since quantum 
effects enhance dispersion, one should expect that quantum solitons will not be so easily found for 
the quantum Zakharov equations as for the classical case~\cite{r26}. We investigate this assumption by taking $[(\partial^2 /\partial \tau^2)(\wtil{\delta n})]=0$ in Eq.~(\ref{q5}) and obtain
\begin{eqnarray}
&&\hspace{-4.8cm}\wtil{\delta n}=-\frac{\mid \wtil{E}\mid^2}{(1-v_0^2)}-\frac{H^2}{(1-v_0^2)^2}\frac{\partial^2}{\partial
X^2}|{\wtil{E}}|^2+O(H^4). \label{q6}
\end{eqnarray}
The detailed calculation of Eq.~(\ref{q6}) is given in Appendix A.

Substituting Eq.~(\ref{q6}) into Eq.~(\ref{q4}), and
keeping only terms of order up to $H^2$ (note that our approximate analysis is only valid for `weakly quantum' plasmas with small $H$), yields the decoupled equation for the envelope $\wtil{E}$ of the modulated Langmuir wave packet:
\begin{eqnarray}
&&\hspace{-3.8cm}\left(i \frac{\partial}{\partial
\tau}+\frac{\partial^2}{\partial X^2}+\frac{|{\wtil{E}}|^2}{(1-v_0^2)}\right)\wtil{E}\approx H^2\frac{\partial^4}{\partial
X^4}\wtil{E}-\frac{H^2}{(1-v_0^2)^2}\wtil{E}\frac{\partial^2}{\partial X^2}|
{\wtil{E}}|^2.\label{q7}
\end{eqnarray}
Eq.~(\ref{q7}) is the quantum-corrected nonlinear Schrodinger equation which is derivable from a variational principle,
\begin{eqnarray}
&&\hspace{-4.8cm}\delta S=\delta\int L dX d\tau=0, \label{q8}
\end{eqnarray}
with the Lagrangian~\cite{r29}
\begin{eqnarray}
&&\hspace{-1.8cm}L=\frac{i}{2}\left(\wtil{E}\frac{\partial \wtil{E}^\ast}{\partial
\tau}-\wtil{E}^\ast\frac{\partial \wtil{E}}{\partial \tau}\right)+\frac{\partial \wtil{E}}{\partial
X}\frac{\partial \wtil{E}^\ast}{\partial X}-\frac{|{\wtil{E}}|^4}{2(1-v_0^2)}+H^2\left[\frac{\partial^2 \wtil{E}}{\partial
X^2}\frac{\partial^2 \wtil{E}^\ast}{\partial
X^2}-\frac{|{\wtil{E}}|^2}{2(1-v_0^2)^2}\frac{\partial^2|{\wtil{E}}|^2}{\partial X^2}\right].\label{q9}
\end{eqnarray}
The variational derivatives $\delta S/\delta \wtil{E}^\ast = \delta S/\delta \wtil{E} = 0$ produce Eq.~(\ref{q7}) and its complex conjugate equation, respectively. The detailed calculation showing this is given in Appendix B.

In this section, we investigate the existence of quantum-modified Langmuir envelope solitons described by the quantum-corrected Zakharov equations. We start by proposing time-dependent
singular solution~\cite{r6} of the form:
\begin{eqnarray}
&&\hspace{-1.8cm}\wtil{E}=\alpha(\tau)\exp\bigg\{i\bigg[\frac{v_0}{2}X+\theta(\tau)\bigg]\bigg\}
\textrm{sech}[\beta(\tau)(X-v_0\tau)],\label{q10}
\end{eqnarray}
where $\alpha$, $\beta$, and $\theta$ are considered as real-valued functions of time only. 
Below we use the variational principle Eq.~(\ref{q8}), requiring that the solution~(\ref{q10}) with arbitrary $\alpha$, $\beta$ and $\theta$ minimize the action, to derive the quantum-corrected form of the coefficients (i. e. $\alpha$, $\beta$, and $\theta$) of soliton solution ~(\ref{q10}).

Substituting Eq.~(\ref{q10}) into Eq.~(\ref{q9}), we get a mechanical system governed by the action
\begin{eqnarray}
&&\hspace{-4.8cm} S=\int L'(\alpha(\tau),\beta(\tau),\theta(\tau)) d\tau, \label{q11}
\end{eqnarray}
where
\begin{eqnarray}
&&\hspace{-4.8cm} L'(\alpha(\tau),\beta(\tau),\theta(\tau))=\int L dX. \label{qq11}
\end{eqnarray}
Now
\begin{eqnarray}
&&\hspace{-1.2cm}~~~~L'(\alpha, \beta, \theta)\nonumber\\
&&\hspace{-1.2cm}~~~=\frac{2\alpha^2\dot\theta}{\beta}+\frac{\alpha^2v_0^2}{2\beta}+\frac{2\alpha^2\beta}
{3}-\frac{2\alpha^4}
{3\beta(1-v_0^2)}+\frac{H^2\alpha^2 v_0^2}{8\beta}+H^2\alpha^2\beta v_0^4+\frac{8H^2\alpha^4\beta}{15(1-v_0^2)^2},\label{q12}
\end{eqnarray}
where $\dot{\theta}=d\theta/d\tau$.

The variational principle Eq.~(\ref{q8}) requires that the variational derivatives $\delta S/\delta\theta=\delta S/\delta\alpha = \delta S/\delta\beta= 0$. Calculating the variational derivatives, we obtain:
\begin{eqnarray}
&&\hspace{-10.8cm}\frac{\partial L'}{\partial
\theta}=0\Rightarrow\frac{d}{d\tau}\left(\frac{\alpha^2}{\beta}\right)=0.\label{q13}
\end{eqnarray}

${\partial L'}/{\partial \alpha}=0$
\begin{eqnarray}
&&\hspace{-1.1cm}\Rightarrow
\dot{\theta}+\frac{v_0^2}{4}+\frac{\beta^2}{3}-\frac{2\alpha^2}
{3(1-v_0^2)}+\frac{7}{15}H^2\beta^4+\frac{1}{2}H^2\beta^2+\frac{1}{16}H^2v_0^4+\frac{8\alpha^2\beta^2H^2}{15(1-v_0^2)^2}=0.\label{q15}
\end{eqnarray}

${\partial L'}/{\partial \beta}=0$
\begin{eqnarray}
&&\hspace{-1.1cm}\Rightarrow
\dot{\theta}+\frac{v_0^2}{4}-\frac{\beta^2}{3}-\frac{\alpha^2}
{3(1-v_0^2)}-\frac{7}{5}H^2\beta^4-\frac{1}{2}H^2\beta^2+\frac{1}{16}H^2v_0^4-\frac{4\alpha^2\beta^2H^2}{15(1-v_0^2)^2}=0.\label{q17}
\end{eqnarray}

Solving Eqs.~(\ref{q13}) - ~(\ref{q17}) for $\alpha$, $\beta$ and $\theta$ enables us to construct the quantum-modified soliton solution of the form Eq.~(\ref{q10}) that is a true solution of Eq.~(\ref{q7}), as it minimizes the action of the system. 
From Eq.~(\ref{q13}) we have
\begin{eqnarray}
&&\hspace{-10.8cm}\dot\alpha=\frac{\alpha}{2}\frac{\dot\beta}{\beta}. \label{qq13}
\end{eqnarray}
Differentiating Eq.~(\ref{q15}) and Eq.~(\ref{q17}) with respect to time and then subtracting, we have
\begin{eqnarray}
&&\hspace{-1.0cm}\frac{4\beta\dot\beta}{3}-\frac{2\alpha\dot\alpha}
{3(1-v_0^2)}+\frac{112}{15}H^2\beta^3\dot\beta+2H^2\beta\dot\beta v_0^2+\frac{12}{15(1-v_0^2)^2}(2\alpha\dot\alpha\beta^2H^2+2\beta\dot\beta\alpha^2H^2)=0.
\label{b18}
\end{eqnarray}
Now substituting the value of $\dot\alpha$ Eq.~(\ref{qq13}) in Eq.~(\ref{b18}) we obtain
\begin{eqnarray}
&&\hspace{-1.0cm}\dot\beta\bigg[\frac{4\beta}{3}-\frac{\alpha^2}
{3\beta(1-v_0^2)}+\frac{112H^2\beta^3}{15}+2H^2\beta v_0^2+\frac{12}{15(1-v_0^2)^2}(2\alpha^2\beta H^2+2\alpha^2\beta H^2)\bigg]=0.
\label{b19}
\end{eqnarray}
From Eq.~(\ref{b19}) we have the following two cases

1) $\dot\beta=0$, $\beta$ is constant.

2) $\big[{4\beta}/{3}-{\alpha^2}/
{3\beta(1-v_0^2)}+{112H^2\beta^3}/{15}+2H^2\beta v_0^2+{12}/{15(1-v_0^2)^2}(2\alpha^2\beta H^2+2\alpha^2\beta H^2)\big]=0$, $\dot\beta$ is constant.

We consider case 1 which implies that $\beta$ is a constant. Then from Eq.~(\ref{q13}), it follows that $\alpha$ is a constant as well. If we consider case 2 it leads to a contradiction with Eq.~(\ref{q13}), since  Eq.~(\ref{q13}) implies that $\alpha^2\propto\beta$ whereas Eq.~(\ref{b19}) implies $\alpha\propto\beta$ (i. e. in classical case), which can only be resolved if $\beta$ is a constant. Therefore, case 1 is true.

In this section we derive expressions for these constants $\alpha$ (amplitude of the soliton), $\beta$ (inverse width), and the corresponding $\theta$ (phase shift) from Eqs.~(\ref{q15}) - ~(\ref{q17}). 

Subtracting Eq.~(\ref{q17}) from Eq.~(\ref{q15}) we have
\begin{eqnarray}
&&\hspace{-2.0cm}\frac{2\beta^2}{3}-\frac{\alpha^2}
{3(1-v_0^2)}+\frac{28}{15}H^2\beta^4+H^2\beta^2v_0^2+\frac{12}{15}\frac{\alpha^2\beta^2H^2}{(1-v_0^2)^2}=0.\label{q18}
\end{eqnarray}
From Eq.~(\ref{q18}), we obtain the quantum-corrected amplitude of the soliton solution~(\ref{q10}) in terms of $\beta$ and $v_0$ in the following form
\begin{eqnarray}
&&\hspace{-1.8cm}\alpha=\sqrt{2(1-v_0^2)}\beta\left[1+\frac{H^2}{20(1-v)^2}(15v_0^2-15v_0^4+\beta^2[52-28v_0^2])\right]+O(H^4).\label{q23}
\end{eqnarray}
Substituting the value of $\alpha$~(\ref{q23}) into Eq.~(\ref{q17}) and integrating over time, we obtain the quantum-corrected phase shift of the soliton solution~(\ref{q10}) in terms of $\beta$ and $v_0$ in the following form
\begin{eqnarray}
&&\hspace{-1.8cm}~~~~{\theta}(\tau)=\bigg[\bigg(\beta^2-\frac{v_0^2}{4}\bigg)-\bigg(\frac{v_0^4}{16}-\frac{\beta^2\{18v_0^2(1-v_0^2)+\beta^2(81-49v_0^2)\}}{15(1-v_0^2)}\bigg)H^2\bigg]\tau+O(H^4).\label{q24}
\end{eqnarray}
Thus, we have $\alpha$~(\ref{q23}) and $\theta(\tau)$~(\ref{q24}) which are defined in terms of $\beta$ and $v_0$. On the other hand, the soliton solution depends on two important parameters, namely, amplitude $E_0$ and the speed $v_0$. Now we express $\alpha$, $\beta$, and $\theta(\tau)$ in terms of $E_0$ and $v_0$ instead of $\beta$ and $v_0$. From Eq.~(\ref{q23}) we obtain the equation for $\beta$ in terms of $E_0$ and $v_0$ as
\begin{figure}[h!]
\begin{center}
\includegraphics[height=90mm,width=100mm]{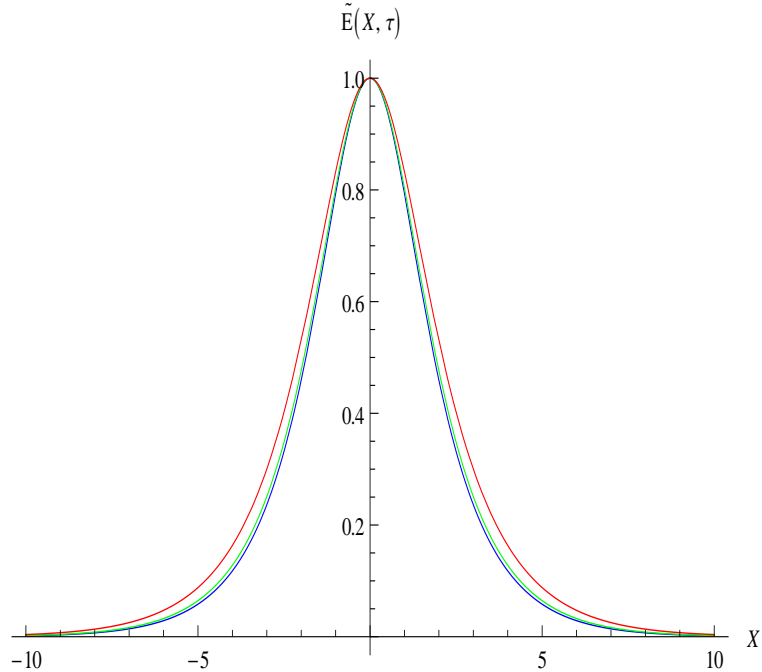}
\caption{The formation of soliton profile in the quantum plasma for $\tau=5$, $v_0=0$, 
$H =0$, (blue line), $H =0.15$, (green line), and $H =0.3$, (red line).}
\end{center}
\end{figure}
\begin{eqnarray}
&&\hspace{-1.1cm}\beta+\frac{H^2\beta}{20(1-v_0^2)}(15v_0^2-15v_0^4+\beta^2[52-28v_0^2])=\frac{E_0}{\sqrt{2(1-v_0^2)}}.\label{q27}
\end{eqnarray}
In the formal classical limit $H=0$ in Eq.~(\ref{q27}), we obtain $\beta=E_0/\sqrt{2(1-v_0^2)}$, which reproduces the classical one-soliton
solution Eq.~(\ref{q10}) with $\alpha=E_0$, $\beta=E_0/\sqrt{2(1-v_0^2)}$, and $\theta=E^2_0/2(1-v_0^2)-v_0^2/4$.
In the quantum case $H>0$ (note that typically $H<<1$ in quantum plasmas at high densities) Eq.~(\ref{q27}) has the perturbative solution
\begin{eqnarray}
&&\hspace{-1.8cm}\beta=\frac{E_0}{\sqrt{2(1-v_0^2)}}-\frac{H^2E_0(15[1-v_0^2]-2E_0[13-7v_0^2])}{20\sqrt{2}(1-v_0^2)^{5/2}}+O(H^4).\label{q28}
\end{eqnarray}
Eq.~(\ref{q28}) represents the quantum-corrected inverse width of the soliton solution~(\ref{q10}) in terms of $E_0$ and $v_0$.

Substituting Eq.~(\ref{q28}) into Eq.~(\ref{q24}), we obtain the quantum-corrected phase shift of the soliton solution~(\ref{q10}) in terms of $E_0$ and $v_0$ in the following form
\begin{eqnarray}
&&\hspace{-1.8cm}\theta(\tau)=\bigg[\frac{E_0^2}{2(1-v_0^2)}\bigg(1+\frac{H^2E_0^2}
{10(1-v_0^2)}\bigg)-\frac{v_0^2}{4}-\frac{H^2v_0^4}{16}\bigg]\tau
+O(H^4),\label{q29}
\end{eqnarray}
or
\begin{eqnarray}
&&\hspace{-12.1cm}~~~~~~~~~~\theta(\tau)=\Omega\tau,\label{q30}
\end{eqnarray}
where
\begin{eqnarray}
&&\hspace{-1.8cm}\Omega=\frac{E_0^2}{2(1-v_0^2)}\bigg(1+\frac{H^2E_0^2}{10(1-v_0^2)}\bigg)-\frac{v_0^2}{4}-\frac{H^2v_0^4}{16}
+O(H^4).\label{q31}
\end{eqnarray}
We used the variational principle Eq.~(\ref{q8}), requiring that the solution~(\ref{q10}) with arbitrary $\alpha$, $\beta$ and $\theta$ minimize the action, to derive the quantum-corrected form of the coefficients (i. e. $\alpha$, $\beta$, and $\theta$) of the soliton solution~(\ref{q10}). Finally, we obtained the expressions for the constants $\alpha$ (amplitude), $\beta$ (inverse width) and the corresponding $\theta$ (phase shift) of the soliton solution~(\ref{q10}) that are expressed in terms of the two arbitrary parameters $v_0$ and $E_0$ in quantum-corrected form (i.e., $\beta$ and $\theta$ in Eqs.~(\ref{q28}) and~(\ref{q29})); note that by definition, $\alpha=E_0$. When $H=0$ in Eqs.~(\ref{q28}) and~(\ref{q29}) then the soliton solution~(\ref{q10}) reduces to the exact soliton solution of the classical NLS. Thus, $\alpha$, $\beta$ and $\theta$ in Eqs.~(\ref{q28}) and~(\ref{q29}), respectively, enable us to construct the quantum-modified soliton solution of the form Eq.~(\ref{q10}) that is a true solution of Eq.~(\ref{q7}), minimizing the action of the system.

We use the quantum-modified soliton solution Eq.~(\ref{q10}) with $\alpha$ , $\beta$ and $\theta$ defined by Eqs.~(\ref{q28}) and ~(\ref{q29}), respectively, to examine cases with different values for $H$ while keeping $E_0$ and $v_0$ fixed. We plot these different cases in Fig.1. It is clear from Fig.1 that for the same amplitude (i.e., $\alpha=E_0$, which is same for both quantum and classical case) the inverse width of the soliton decreases with quantum correction terms proportional to $H^2$; i.e., the width increases with quantum-correction term $H$. In other words, the quantum effects lead to widening of the soliton in quantum plasma, compared to the classical soliton of the same amplitude.

\section{Conclusion}
We have investigated the quantum-corrected Zakharov equations and the
existence of quantum-corrected solitons in a fully nonlinear quantum
plasma. We constructed a solution using a variational principle (requiring that the solution minimizes the action of the system) and also derived the quantum-modified coefficients of the soliton solution of the quantum-modified NLS equation. We found that for the same amplitude (i.e., $\alpha=E_0$, for both quantum and classical cases) the quantum effects change the inverse width $(\beta)$ and the phase shift $(\theta)$ given by Eqs.~(\ref{q28}) and~(\ref{q29}) for the soliton solution Eq.~(\ref{q10}). In Fig.1 we have shown that for the same amplitude (i.e., $\alpha=E_0$, is same for both quantum and classical case) the inverse width of the soliton decreases with quantum correction terms proportional to $H^2$; i.e., the width increases with the quantum-correction term $H$. We will study in a subsequent paper how the quantum correction terms proportional to $H^2$ in Eqs.~(\ref{q28}),~(\ref{q29}) affect the stability of such solutions. In particular, we will numerically study the ultimate fate of the solitons in the quantum plasmas and will also analyse the collision of two solitons for both cases where $H=0$ and $H>0$.

\appendix
\section{}
The dimensionless Zakharov equations in quantum plasma can be presented in the form
\begin{eqnarray}
&&\hspace{-3.8cm}\left(i \frac{\partial}{\partial
\tau}+\frac{\partial^2}{\partial X^2}-H^2\frac{\partial^4}{\partial
X^4}\right)\wtil{E}=\wtil{\delta n} \wtil{E},\label{a1}
\end{eqnarray}
\begin{eqnarray}
&&\hspace{-2.8cm}\left(\frac{\partial^2}{\partial
\tau^2}+H^2\frac{\partial^4}{\partial \tau^2
\partial X^2}
-\frac{\partial^2}{\partial X^2}\right)\wtil{\delta n}
=\frac{\partial^2}{\partial
X^2}\left(1+H^2\frac{\partial^2}{\partial X^2}\right){|
\wtil{E}|^2}.\label{a2}
\end{eqnarray}
The solution of these nonlinear equations is obtained by transforming the independent variables, 
using
\begin{eqnarray}
&&\hspace{-2.8cm}\eta=X-v_0\tau,\label{a3}
\end{eqnarray}
where $v_0$ is a constant speed at which the soliton solution propagates.

Differentiating Eq.~(\ref{a3}) with respect to $\tau$ and $X$ respectively, we obtain
\begin{eqnarray}
&&\hspace{-2.8cm}\frac{\partial}{\partial\tau}=-v_0\frac{\partial}{\partial\eta},\label{a4}
\end{eqnarray}
\begin{eqnarray}
&&\hspace{-2.8cm}\frac{\partial}{\partial X}=\frac{\partial}{\partial\eta}.\label{a5}
\end{eqnarray}

Substituting the values of Eq.~(\ref{a4}) and Eq.~(\ref{a5}) in Eq.~(\ref{a2}), we obtain the 
following equation

\begin{eqnarray}
&&\hspace{-2.8cm}(v_0^2-1)\wtil{\delta n}+H^2v_0^2\frac{\partial^2}{\partial \eta^2
}\wtil{\delta} n
=|\wtil{E}|^2+H^2\frac{\partial^2}{\partial \eta^2
}|\wtil{E}|^2.\label{a6}
\end{eqnarray}
The iterative solution of Eq.~(\ref{a6}) is as follows:

When $H=0$ then Eq.~(\ref{a6}) becomes

\begin{eqnarray}
&&\hspace{-2.8cm}\wtil{\delta n}^{(1)}
=-\frac{|\wtil{E}|^2}{(1-v_0^2)}.\label{a7}
\end{eqnarray}
When $H\neq0$ then Eq.~(\ref{a6}) can be written as

\begin{eqnarray}
&&\hspace{-2.8cm}H^2v_0^2\frac{\partial^2}{\partial \eta^2
}\wtil{\delta n}^{(1)}+{(v_0^2-1)}\wtil{\delta} n^{(2)}
=|\wtil{E}|^2+H^2\frac{\partial^2}{\partial \eta^2
}|\wtil{E}|^2,\label{a8}
\end{eqnarray}
so that
\begin{eqnarray}
&&\hspace{-2.8cm}\wtil{\delta} n^{(2)}
=\wtil{\delta n}^{(1)}-\frac{H^2}{(1-v_0^2)}\frac{\partial^2}{\partial \eta^2
}|\wtil{E}|^2-\frac{H^2v_0^2}{(1-v_0^2)}\frac{\partial^2}{\partial \eta^2
}\wtil{\delta n}^{(1)}.\label{a9}
\end{eqnarray}
Substituting Eq.~(\ref{a7}) into Eq.~(\ref{a9}) we obtain
\begin{eqnarray}
&&\hspace{-2.8cm}\wtil{\delta n}^{(2)}=
-\frac{|\wtil{E}|^2}{(1-v_0^2)}-\frac{H^2}{(1-v_0^2)^2}\frac{\partial^2}{\partial \eta^2
}|\wtil{E}|^2.\label{a10}
\end{eqnarray}
again
%\begin{eqnarray}
%&&\hspace{-2.8cm}H^2v_0^2\frac{\partial^2}{\partial \eta^2
%}\wtil{\delta n}^{(2)}+{(v_0^2-1)}\wtil{\delta} n^{(3)}
%=|\wtil{E}|^2+H^2\frac{\partial^2}{\partial \eta^2
%}|\wtil{E}|^2,\label{a10}
%\end{eqnarray}
\begin{eqnarray}
&&\hspace{-2.8cm}\wtil{\delta n}^{(3)}
=\wtil{\delta n}^{(1)}-\frac{H^2}{(1-v_0^2)}\frac{\partial^2}{\partial \eta^2
}|\wtil{E}|^2-\frac{H^2v_0^2}{(1-v_0^2)}\frac{\partial^2}{\partial \eta^2
}\wtil{\delta n}^{(1)}+O(H^4).\label{a11}
\end{eqnarray}
Finally, we can write
\begin{eqnarray}
&&\hspace{-2.8cm}\wtil{\delta n}
=-\frac{|\wtil{E}|^2}{(1-v_0^2)}-\frac{H^2}{(1-v_0^2)^2}\frac{\partial^2}{\partial X^2
}|\wtil{E}|^2+O(H^4).\label{a12}
\end{eqnarray}
\section{}
The quantum-corrected nonlinear Schrodinger equation can be written as
\begin{eqnarray}
&&\hspace{-3.8cm}\left(i \frac{\partial}{\partial
\tau}+\frac{\partial^2}{\partial X^2}+\frac{|{\wtil{E}}|^2}{(1-v_0^2)}\right)\wtil{E}\approx H^2\frac{\partial^4}{\partial
X^4}\wtil{E}-\frac{H^2}{(1-v_0^2)^2}\wtil{E}\frac{\partial^2}{\partial X^2}|
{\wtil{E}}|^2,\label{b1}
\end{eqnarray}
which is derivable from a variational principle,
\begin{eqnarray}
&&\hspace{-4.8cm}\delta S=\delta\int L dX d\tau=0, \label{b2}
\end{eqnarray}
based on the Lagrangian~\cite{r29}
\begin{eqnarray}
&&\hspace{-1.8cm}L=\frac{i}{2}\left(\wtil{E}\frac{\partial \wtil{E}^\ast}{\partial
\tau}-\wtil{E}^\ast\frac{\partial \wtil{E}}{\partial \tau}\right)+\frac{\partial \wtil{E}}{\partial
X}\frac{\partial \wtil{E}^\ast}{\partial X}-\frac{|{\wtil{E}}|^4}{2(1-v_0^2)}+H^2\left[\frac{\partial^2 \wtil{E}}{\partial
X^2}\frac{\partial^2 \wtil{E}^\ast}{\partial
X^2}-\frac{|{\wtil{E}}|^2}{2(1-v_0^2)^2}\frac{\partial^2|{\wtil{E}}|^2}{\partial X^2}\right],\label{b3}
\end{eqnarray}
where $L=L(\wtil{E}, \wtil{E}^\ast, \partial \wtil{E}/\partial X, \partial \wtil{E}^\ast/\partial X,t)$.
The variational derivatives $\delta S/\delta \wtil{E}^\ast = \delta S/\delta \wtil{E} = 0$ produce Eq.~(\ref{b1}) and its complex conjugate equation, respectively.
The corresponding Lagrange equation of the quantum-corrected nonlinear Schrodinger equation can be written as
\begin{eqnarray}
&&\hspace{-1.8cm}\frac{\partial}{\partial
\tau}\frac{\partial L}{\partial \left(\frac{\partial \wtil{E}^\ast}{\partial
\tau}\right)}+\frac{\partial}{\partial
\ X}\frac{\partial L}{\partial \left(\frac{\partial \wtil{E}^\ast}{\partial
X}\right)}-\frac{\partial L}{\partial \wtil{E}^\ast}=0.\label{b4}
\end{eqnarray}
From Eq.~(\ref{b3}), we obtain
\begin{eqnarray}
&&\hspace{-1.8cm}\frac{\partial L}{\partial \left(\frac{\partial \wtil{E}^\ast}{\partial
X}\right)}=\frac{\partial \wtil{E}}{\partial X}-H^2\frac{\partial^3 \wtil{E}}{\partial
X^3},\label{b6}
\end{eqnarray}
\begin{eqnarray}
&&\hspace{-1.8cm}\frac{\partial L}{\partial \left(\frac{\partial \wtil{E}^\ast}{\partial
\tau}\right)}=\frac{i}{2}\wtil{E},\label{b7}
\end{eqnarray}
and
\begin{eqnarray}
&&\hspace{-1.8cm}\frac{\partial L}{\partial
\wtil{E}^\ast}=-\frac{i}{2}\frac{\partial \wtil{E}}{\partial \tau}-\wtil{E}|\wtil{E}|^2-H^2\wtil{E}\frac{\partial^2 |\wtil{E}|^2}{\partial X^2}.\label{b8}
\end{eqnarray}

Substituting the values of Eqs.~(\ref{b6}) -~(\ref{b8}) into Eq.~(\ref{b4}) one can obtain the quantum-corrected nonlinear Schrodinger equation Eq.~(\ref{b1}).
\\\\
\noindent{\bf Acknowledgments.} This study was partially supported by the 
Australian Research Council (ARC) and by Asian Office 
of Aerospace R and D under grant number FA2386-12-1-4077 and JSPS Grant-in-Aid for 
Scientific Research (A) under Grant 23244110.

\end{sloppypar}
\end{document}